# THE EVOLUTION OF THE OPTICAL GALAXY LUMINOSITY FUNCTION

MATTHEW COLLESS

*Mount Stromlo & Siding Spring Observatories*
*The Australian National University*
*Canberra, ACT 2611, Australia*

## ABSTRACT

The optical luminosity function is a fundamental characterization of the galaxy population. A combination of earlier redshift surveys with two new surveys allows the first accurate determination of the evolution of the luminosity function with redshift, and reveals a marked steepening of the faint end slope. This effect is more profound for star-forming galaxies—there are 5–10 times as many star-forming galaxies at $z\sim0.5$ as there are locally. These results, together with high-resolution imaging and linewidth velocity measurements, support the view that the excess of star-forming galaxies at moderate redshift represents a general increase in the star-formation rate of normal galaxies rather than a distinct new population. This increase in star-formation appears in lower-$L$ galaxies at lower redshift and only appears in $L^*$ galaxies at $z\gtrsim0.5$. Imaging studies provide indirect evidence which suggests that interactions are responsible for a large part of this increased activity.

## 1. Introduction

For some time now, the evidence from number counts of faint galaxies and deep redshift surveys has pointed towards some form of evolution in the galaxy population at redshifts $z<0.5$. There is now good agreement in the observed number counts[18] out to $B\sim27$ and $K\sim20$, and in the redshift distribution[11] out to $B\sim24$. Despite some uncertainties in characterizing the local galaxy population[12] and the different selection biases of bright and faint galaxy samples, these observations make it hard to escape the conclusion[1,5] that the optical galaxy luminosity function (LF) has evolved significantly since quite modest redshifts. Comparisons with models for the form of this evolution rule out the simplest luminosity evolution picture but are unable to distinguish between the other scenarios, such as an increased merger rate or a population of fading dwarf galaxies.

This paper begins by discussing the uncertainties concerning the local (i.e. $z\sim0$) LF, and the various problems encountered in attempting to determine the extent of its evolution from earlier redshift surveys. The main body of the paper reports the first accurate determination of the LF as a function of redshift, resulting from the combination of previous studies and two new redshift surveys. With a surer grasp of the form of the LF evolution, we then turn to attempts to reveal the physical processes that drive it. We outline results from high-resolution multicolour imaging concerning the morphology of the star-forming galaxies and the role of interactions, and report some preliminary measurements of the linewidth velocities in these objects. The final section sums up the evidence to date on galaxy evolution.

## 2. Problems in Luminosity Function Determinations

*2.1. The Local Luminosity Function*

Our detailed knowledge of the local LF is based on redshift surveys of bright ($B$<17) galaxies. These surveys[9,14,16] suffer from two main deficiencies that principally affect the determination of the faint-end slope of the LF: (i) the bright apparent magnitude limits of these surveys correspond to relatively small sample volumes for low-luminosity galaxies, with consequent vulnerability to local clustering; (ii) the high surface brightness thresholds used to detect galaxies in these surveys will exclude any population of low surface brightness galaxies[17]. The relatively small numbers of galaxies in these redshift surveys also limits the extent to which the local LF can be determined for each spectral or morphological type.

These uncertainties in the faint end of the LF may lie behind the embarrassing difference between the observed and predicted number counts for *bright* galaxies[15]—a no-evolution model using the local determinations of the LF severely under-predicts the slope of the number counts brighter than $B$=19. Other possible explanations for this difference are that the bright number counts may be incorrect, due perhaps to magnitude scale errors[19]; or that there is very large scale structure modulating the number counts; or that there is strong evolution of the LF even at redshifts as low as $z$∼0.1.

All the problems with determining the local luminosity function can be overcome with sufficiently large surveys using suitable selection criteria. However until this is done the local LF remains a shaky foundation on which to build a picture of galaxy evolution.

*2.2. The Luminosity Function at Higher Redshifts*

Apart from a few preliminary attempts[8,13], it has not been considered worthwhile until now to attempt to directly determine the galaxy LF for $z$>0. Instead the preferred method for investigating galaxy evolution has been through a comparison of the observed number counts and redshift distributions with the predictions of various models[1]. This approach has the merit of leaving the data in as raw a state as possible and putting the onus on the modeller to correctly simulate the detailed observational filter through which the galaxies pass. The few-parameter evolutionary models employed were also well-matched to the relatively small samples from the redshift surveys.

However with a better understanding of the observational filters (i.e. the selection as a function of surface brightness, size and so on) and with larger samples of galaxies, this approach becomes too restrictive and it is better to directly determine the LF as a function of redshift. This requires moving away from simple surveys with a single magnitude limit, since these inevitably result in samples of galaxies in which the luminosity of the galaxies is strongly correlated with their redshift. The LFs determined in different redshift ranges from such a survey will each cover a relatively narrow luminosity range, with little overlap in luminosity between one redshift range and the next.

In the next section we report a study based on a combination of old and new redshift surveys which remedies some of the above problems and yields a direct determination of the evolution in the optical galaxy luminosity function out to $z$~1.

## 3. The Redshift Surveys

### 3.1. Observational Parameters

In order to cover the largest possible range in apparent magnitude (and hence a wide range in luminosity at each redshift), we have combined all the various redshift surveys made by our group over several years with two new redshift surveys. The first of these new surveys[10] was carried out with the Autofib multifibre spectrograph on the Anglo-Australian Telescope and resulted in 1026 redshifts for galaxies with $b_J$=17–22; the second[11] used the LDSS-2 multislit spectrograph on the William Herschel Telescope to obtain redshifts for 73 galaxies with $b_J$=22.5–24. When combined with the earlier DARS[20], BES[2] and LDSS-1[4,5] redshift surveys, the net result is a sample of more than 1700 redshifts spanning the magnitude range $b_J$=11.5–24. The following table gives the magnitude range, area, number of fields and completeness of each survey.

Table 1 – The Redshift Surveys

| Survey | $b_J$ range | Area (□°) | # Fields | Complete-ness | # Gals |
|---:|:---:|:---:|:---:|:---:|---:|
| DARS | 11.5–16.8 | 70.8 | 5 | 96% | 326 |
| BES | 20.0–21.5 | 0.50 | 5 | 83% | 188 |
| LDSS-1 | 21.0–22.5 | 0.12 | 6 | 82% | 100 |
| Autofib bright | 17.0–20.0 | 5.5 | 16 | 70% | 480 |
| Autofib faint | 19.5–22.0 | 4.7 | 32 | 81% | 546 |
| LDSS-2 | 22.5–24.0 | 0.07 | 5 | 71% | 73 |

### 3.2. Spectral Types and K-Corrections

A crucial element in converting the redshift surveys into luminosity functions is obtaining a spectral type for each galaxy in order to derive K-corrections. The main difficulties are that none of the survey spectra are accurately flux-calibrated, and that the fainter surveys generally have modest S/N and are subject to systematic sky-subtraction errors. We cannot therefore simply fit the spectra to a library of templates, but must instead use indirect means to determine spectral types.

For DARS we have morphological types which correlate moderately well with spectral type, and in any case K-corrections are less important for these low-redshift objects. For the LDSS-1 and LDSS-2 surveys we have $b_J-r_F$ or $B-R$ colours which can readily, and with sufficient accuracy, be mapped to spectral type. However for the Autofib and BES surveys we have no colours and must use the spectra themselves.

For these surveys the spectra are typed by a cross-correlation technique very similar to that used in estimating redshifts. The spectra are filtered to remove the continuum and then cross-correlated with template spectra drawn from Kennicutt's library. Simulations show that the spectral type is correctly assigned 80% of time, and to within ±1 type class >90% of the time. Comparisons of types determined this way and from colours also show good agreement. The typing errors increase beyond $z\sim0.5$ due to the smaller spectral overlap with the templates, but there are few galaxies at such redshifts in BES or Autofib surveys.

### 3.3. Incompleteness

With the exception of the essentially-complete bright DARS survey, all the surveys are 15–30% incomplete in redshift identifications. Within each survey the completeness at the bright end is close to 100% but then drops as we go to fainter magnitudes. We argue that this effect is largely due to lower S/N at fainter magnitudes within each survey, in which case it is correctable by weighting so long as it is independent of spectral type and redshift.

To test whether incompleteness depends on spectral type, we have computed the $V/V_{max}$ statistic for each galaxy and then examined the $V/V_{max}$ distribution for each spectral type before and after applying the magnitude-dependent completeness correction. Although the unweighted $V/V_{max}$ distributions depart very significantly from the uniform distributions expected if there were no incompleteness, the weighted distributions are much closer to this ideal, implying that any type-dependence of incompleteness is small compared to the main dependence on apparent magnitude.

We test whether the incompleteness depends on redshift by examining the redshift distributions in the overlap range between the faint (low-completeness) end of one survey and the bright (high-completeness) end of the next-fainter survey. The good agreement we find implies that our incompleteness is not due to missing objects at high (or low) redshift. Of course we cannot test the faintest LDSS-2 survey this way and it may indeed suffer from missing objects at redshifts greater than $z\sim1$ as [OII] and H+K enter sky bands.

## 4. Luminosity Functions

### 4.1. Computing Luminosity Functions

We use both a $1/V_{max}$ estimator and a modification of the step-wise maximum likelihood (SWML) method to derive LFs. The former is a more direct and unbiased estimator, while the latter is more insensitive to clustering. LF errors for $1/V_{max}$ estimates are derived by bootstrap simulations, while the SWML method provides its own error estimates. The error estimates for the two methods are in reasonable agreement. We also compute LFs both with and without the correction for magnitude-dependent incompleteness described in the previous section, in order to see the effect of incompleteness on our results.

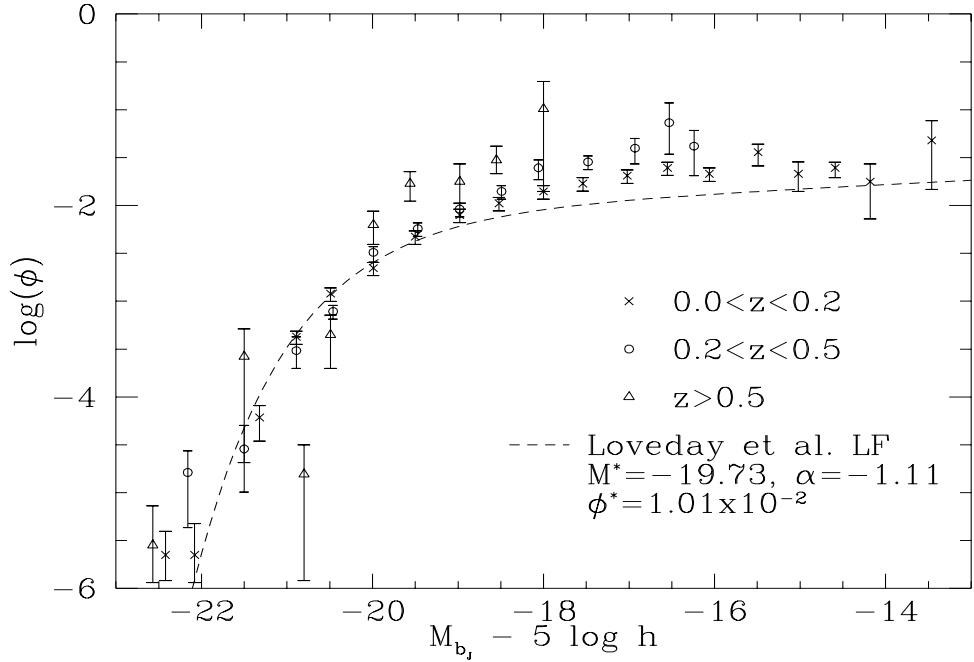

Figure 1: The evolution of the galaxy luminosity function in three redshift bins: $z$=0–0.2, $z$=0.2–0.5 and $z$>0.5. The local luminosity function[14] is shown for comparison.

We find that in fact there is very little difference in the derived LFs from the two estimators, whether with or without the completeness correction, suggesting that the results are robust with respect to estimation algorithm, clustering and incompleteness.

*4.2. Evolution in the Luminosity Function*

The luminosity functions that we derive from our combined sample are shown in Figure 1. The figure shows clear evidence for evolution of the luminosity function in each of the three redshift bins $z$=0.0–0.2, $z$=0.2–0.5 and $z$>0.5. The change in the observed LF looks like a steepening of the faint-end slope (or possibly density evolution). Bright ($\sim L^*$) galaxies show no significant brightening out to $z\sim0.5$, but perhaps some brightening at $z$>0.5. Applying $\chi^2$ one- and two-sample tests shows that: the LF at $z$=0.0–0.2 differs from local LF at a confidence level >99.9%; the $z$=0.2–0.5 LF differs from the $z$=0.0–0.2 LF at the 99.9% level; and the $z$>0.5 LF differs from the $z$=0.2–0.5 LF at the 98% level.

We have applied various tests to check that the observed evolution is not an artifact of some incorrect input. First, we have checked that the results are not due to incorrect K-corrections by computing observed LFs and no-evolution models in the *observer's* frame (i.e. with no K-corrections). In this frame we cannot intercompare LFs for different redshift ranges, but we can compare each observed LF with the corresponding no-evolution model. When we do so, we see the same trends emerge— the faint end of the observed LF becomes progressively steeper than the faint end of the no-evolution LF—leading us to conclude that the evolution we see is not the product of incorrect K-corretions.

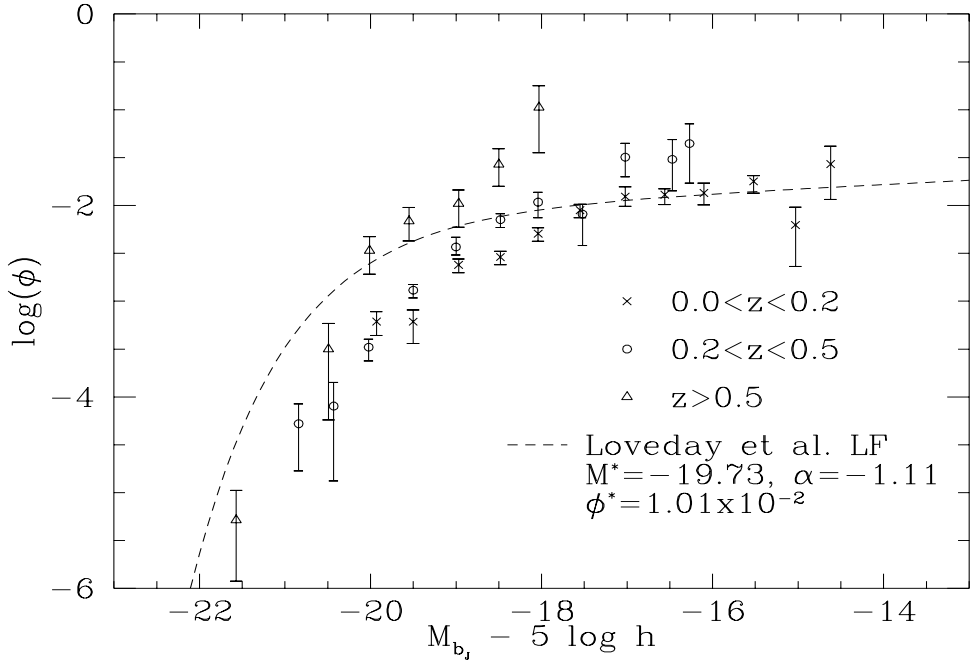

Figure 2: The evolution of the luminosity function for star-forming galaxies in the three redshift bins $z$=0–0.2, $z$=0.2–0.5 and $z$>0.5.

We have also checked that the LFs are not sensitive to the errors we make in assigning spectral types to the galaxies. To do this we have re-computed the LFs after randomly re-assigning spectral types for 20% of galaxies by ±1 spectral type (roughly the extent of the typing errors we find in our simulations—see above). The LFs derived in this manner differ negligibly from the observed LFs, so our results are robust against spectral typing errors.

*4.3. Star-Forming Galaxies*

A $V/V_{max}$ test shows that the galaxies with high [OII] equivalent widths (EW>20Å) form a virtually complete subsample (except possibly at $z$>1). The LFs for these star-forming galaxies (shown in Figure 2) display an evolution that is qualitatively similar to that of the whole sample but even more pronounced. The figure shows that there were many more star-forming sub-$L^*$ galaxies at $z$~0.3 than there are now, and that beyond $z$~0.5 the number of star-forming galaxies has increased by nearly an order of magnitude compared to the present at all luminosities up to and including $L^*$.

Another way of quantifying the evolution of the star-forming population is to consider the median [OII] equivalent width of the galaxy population as a function of redshift and luminosity. This characteristic equivalent width increases with $z$ at all fixed values of $L$, implying higher past rates of star-formation in galaxies of *all* luminosities. The difference between bright and faint galaxies is that bright galaxies begin to show this increase in star-formation rate at higher redshifts than faint galaxies.

## 5. High-Resolution Multicolour Imaging

Although we now have a quantitative measure of the evolution in the galaxy luminosity function we still need further information, as the available observations—essentially counts, colours and redshifts—are insufficient to discriminate between the various models proposed for the origin of the evolution in the galaxy population. Some initial results from two other approaches to discovering the physical mechanisms behind this evolution are discussed in this and the following section.

The first of these approaches is to follow-up the deep redshift surveys with high-resolution multicolour imaging. In principle such imaging can yield surface brightness distributions for the galaxies, their sizes and spatial colour profiles, and whether or not they have distorted morphologies or close companions. Such information would enable us to investigate whether these galaxies are notably compact or of low surface brightness, to locate the sites of star-formation in each galaxy, and to examine whether their star formation might be induced by tidal interactions or mergers with other galaxies.

### 5.1. Morphology at z~0.3

Because the evolution is occurring at relatively low redshifts ($z$<0.5), the best ground-based images (with 0″.5 seeing) can resolve features on scales as small as $\sim 2$ $h^{-1}$ kpc. We have taken advantage of this fact to follow-up the LDSS-1 redshift surveys with an imaging programme[7] using HRCam, a fast-guiding camera with a 2′.2 FoV and 0″.1 pixels on the CFHT.

The galaxies we imaged had $b_J$=21–22.5 and redshifts and [OII] EWs from the LDSS-1 surveys. They fell into two samples: (i) 17 'star-forming' galaxies with EW>20Å, and (ii) 9 'quiescent' galaxies with EW<10Å. We used HRCam to obtain deep 0″.5–1″.0 images in V and I or B and I. The quality of these images is good enough to allow us to say whether the object is dominated by an exponential disk, an $r^{1/4}$ bulge or a point source, and also to determine the appropriate scalelength to $\sim 10\%$ precision via direct $\chi^2$-fitting of a 2D seeing-convolved model for the surface brightness.

For this sample of 0.1<z<0.7 star-forming galaxies, with absolute magnitudes in the range M*−1 to M*+5, we find that the great majority are dominated by exponential disks with scalelengths from 0.3–6 $h^{-1}$ kpc. Figure 3 shows that the size–luminosity relation for this sample is identical to that for low-redshift normal spirals. In general, we find that each individual galaxy at z~0.3 has a colour, [OII] EW, size and luminosity that is consistent with some type of z~0 galaxy. Thus there is no large population at z~0.3 which is not found at z~0, and the observed evolution must be due to a change in the relative fractions of different galaxy types, and not a distinctly different new population.

### 5.2. Location & Origin of Star-Formation

For almost all the star-forming galaxies we imaged, the blue (B or V) scalelength is at least as large as the red (I) scalelength, implying that star-formation is not confined to the nucleus but is occurring broadly across disk. The two exceptions to this rule are galaxies with close companions that are dominated by unresolved blue sources in the nuclear region.

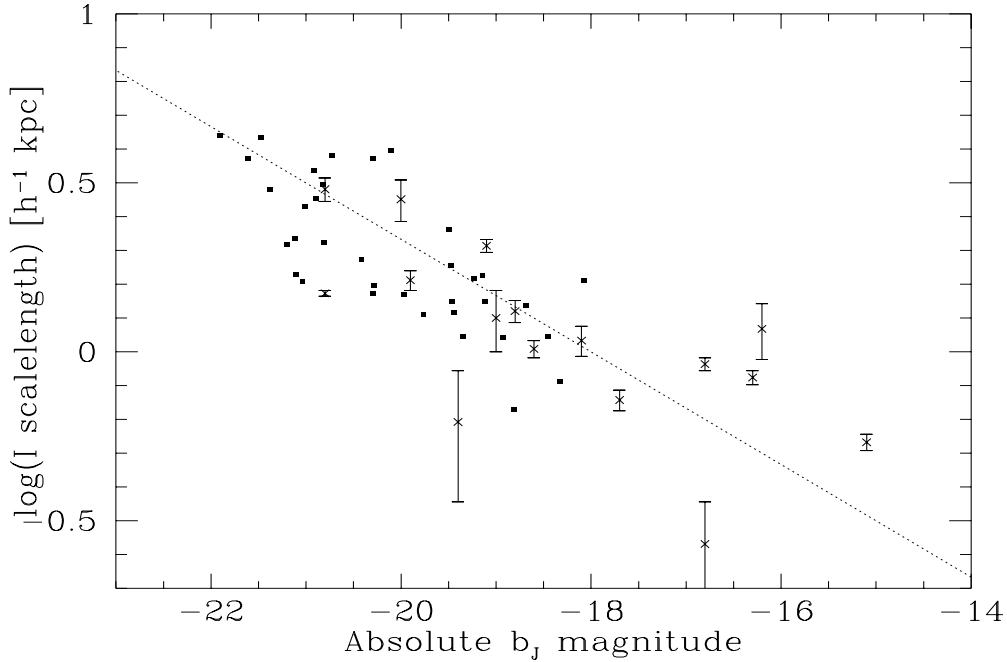

Figure 3: The $I$-band size-luminosity relation for local galaxies (dots) and galaxies at moderate redshift (crosses). The dotted line is a Holmberg relation, $M = -6\log(\text{size}) + C$.

One of the most significant observations to emerge from this study is that 5 of the 17 high-EW galaxies (∼30%) turn out to have companions closer than 10 $h^{-1}$ kpc, while there are no companions to any of the 9 low-EW galaxies (<10%). A $t$-test rejects the hypothesis that the two samples have the same fraction of close companions at the 1% level. The point worth noting is that the fraction of star-forming galaxies with close companions is similar to the fractional excess of high-EW galaxies found in the redshift surveys. This similarity suggests that the excess of star-forming galaxies at moderate redshift may in large part be due to interactions of some sort. However interactions are not the whole story: several of the high-EW galaxies do not have close companions, and there are also three galaxies with very blue colours but low EWs and unresolved morphologies.

## 6. Linewidths for Faint Galaxies

A second approach to investigating the processes behind galaxy evolution at moderate redshifts is the attempt to measure linewidth velocities for faint blue galaxies[6]. From such linewidths one can estimate the masses of the faint blue galaxies, as opposed to their luminosities, and look to see whether they obey (or deviate from) the Tully-Fisher and Faber-Jackson relations. In this way we could distinguish bursts of star-formation in dwarfs from milder evolution in normal late-type spirals, and help

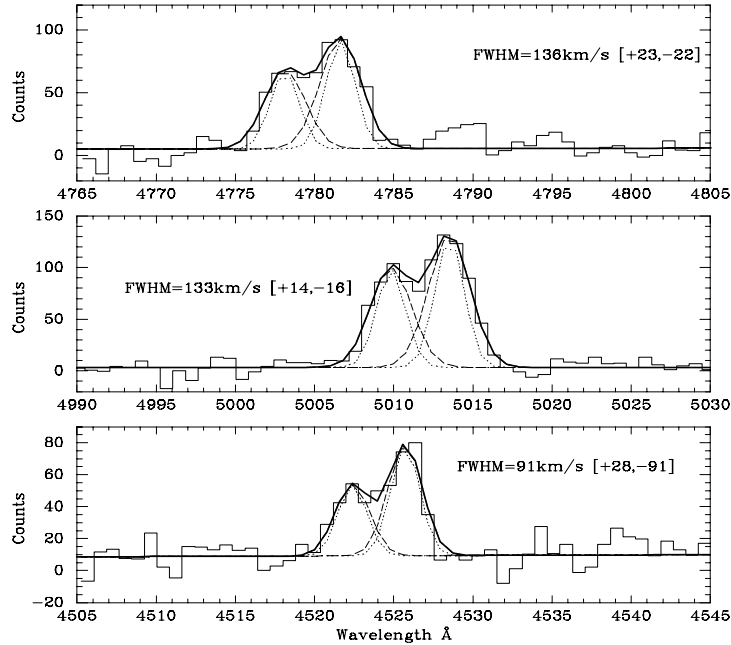

Figure 4: The Gaussian fits to the [OII] line profiles for three star-forming galaxies with $b_J$=21.25–22.0 and $z$=0.15–0.35.

to identify the present-day counterparts of the blue galaxies at intermediate redshifts.

In an initial attempt to follow this programme, we have used the Autofib fibre spectrograph on the AAT to get spectra for a sample of blue ($b_J - r_F < 1.2$) galaxies with $b_J$=21.25–22.0. These spectra cover 800Å in a range centred on 4700Å, corresponding to [OII] 3727Å at $z$=0.15–0.35. In this magnitude range we expect 1-in-6 galaxies to have both $z$=0.15–0.35 and [OII] EW>20Å, and hence have a measurable [OII] linewidth.

Of the 54 galaxies in the one field observed to date, 24 have detectable [OII]. We have fitted a Gaussian broadening profile to the [OII] doublet keeping the separation of the features fixed but allowing the relative strengths to vary. Figure 4 shows examples of the fitted line profiles. Of the 24 detections, 9 have $\sigma < 70\,\mathrm{km\,s^{-1}}$ (below the lower limit of detectable broadening) and 15 had measured $\sigma$'s of 70–200 $\mathrm{km\,s^{-1}}$ (11 with $\sigma > 100\,\mathrm{km\,s^{-1}}$).

Although this work is at an early stage, one preliminary conclusion we can draw is that although some of the faint blue galaxies at $z$=0.15–0.35 may be dwarfs (those with linewidths less than $70\,\mathrm{km\,s^{-1}}$), the majority have velocities (and presumably masses) that are typical of normal present-day galaxies.

## 7. Conclusions

The various studies discussed above now provide us with a wide variety of observations concerning the evolution of the galaxy population at moderate redshifts. The

main results of these studies could be summarized as follows:

1. Out to $z$~0.5 the $B$-band LF shows a progressive steepening of the faint end.

2. At $z \gtrsim 0.5$ $L^*$ galaxies begin to participate, and the overall effect on the LF looks like a combination of both luminosity and density evolution. There is no evidence for luminosity evolution of bright galaxies until $z$~0.5.

3. The star-forming population shows the same overall trends in its LF, but more strongly. There are 5–10 times as many star-forming galaxies at $z \gtrsim 0.5$ as at $z$~0 at all luminosities up to and including $L^*$.

4. The median [OII] EW increases with redshift at all luminosities, implying an increasing mean star-formation rate. This increase begins at higher redshift for brighter galaxies.

5. The correlation function of the star-forming galaxies brighter than $b_J$=22 is indistinguishable from that of the quiescent galaxies for separations $>3$ h$^{-1}$ Mpc. Thus at $z$~0.3 the star-forming galaxies are clustered in the same way as the rest of the population on large scales[3].

6. On much smaller scales ($\lesssim 10$ h$^{-1}$ kpc), high-resolution imaging shows the star-forming galaxies to have a much higher incidence of close companions, suggesting that interactions are involved in their increased activity.

7. High-resolution imaging also shows that moderate-redshift galaxies have morphologies and a size–luminosity relation very similar to normal $z$~0 galaxies.

8. The majority of $z$~0.3 star-forming galaxies have linewidth velocities comparable to those of normal present-day galaxies (i.e. $\sigma > 100 \, \mathrm{km\,s^{-1}}$), though a sizeable fraction have $\sigma < 70 \, \mathrm{km\,s^{-1}}$.

All these points support the view that the excess of star-forming galaxies at moderate redshift is not a distinct new population, but rather reflects a marked general increase in the star-formation rate of normal galaxies. This effect appears in lower-$L$ galaxies at lower redshift and only appears in $L^*$ galaxies at $z \gtrsim 0.5$. Statistical evidence from the numbers of close companions points to interactions being responsible for a large part of this increased star-formation.

We are currently obtaining HST images in UV and $I$ bands of galaxies in the $B$<24 LDSS-2 redshift survey in order to push these morphological observations to higher redshifts. Other future plans include a redshift survey of $>10^5$ galaxies using the AAT's two-degree field spectrograph (2dF), which will come on-line in 1995. As well as addressing questions of the large-scale structure in the galaxy distribution, such a survey will yield the present-day LF down to very faint absolute magnitudes with surpassing precision and so reveal in detail the LF as a function of local overdensity and spectral/morphological type.

## 8. Acknowledgements

It is a pleasure to thank Tom Broadhurst, Richard Ellis, Karl Glazebrook, Raja Guhathakurta, Jeremy Heyl, Hans-Walter Rix and David Schade for stimulating collaborations and for allowing me to report some unpublished work.

## 9. References

1. Broadhurst T.J., Ellis R.S., Glazebrook K., 1992, *Nature*, **355**, 55.
2. Broadhurst T.J., Ellis R.S., Shanks T., 1988, *MNRAS*, **235**, 837.
3. Cole S., Ellis R.S., Broadhurst T.J. & Colless M.M., 1993, *MNRAS*, **267**, 541.
4. Colless M.M., Ellis R.S., Taylor K. & Hook R.N., 1990, *MNRAS*, **244**, 408.
5. Colless M.M., Ellis R.S., Broadhurst T.J., Taylor K. & Peterson B.A., 1993, *MNRAS*, **261**, 19.
6. Colless M.M., Guhathakurta P., Rix H-W., 1994, in preparation.
7. Colless M.M., Schade D.J., Broadhurst T.J. & Ellis R.S., 1994, *MNRAS*, **267**, 1108.
8. Eales S., 1993, *Ap.J.*, **404**, 51.
9. Efstathiou G.P., Ellis R.S., Peterson B.A., 1988, *MNRAS*, **232**, 431.
10. Ellis R.S., Colless M.M., Broadhurst T.J., Heyl J.S., Glazebrook K., 1994, in preparation.
11. Glazebrook K., Ellis R.S., Colless M.M., Broadhurst T.J., Allington-Smith J.R., Tanvir N.R. & Taylor K., 1994, *MNRAS*, in press.
12. Koo D.C., Kron R.G., 1992, *Ann.Rev.Astron.Astrophys.*, **30**, 613.
13. Lonsdale C.J., Chokshi A., 1993, *A.J.*, **105**, 1333.
14. Loveday J., Peterson B.A., Efstathiou G.P., Maddox S.J., 1992, *Ap.J.*, **390**, 338.
15. Maddox S.J., Sutherland W.J., Efstathiou G.P., Loveday J., Peterson B.A., 1990, *MNRAS*, **247**, 1P.
16. Marzke R.O., Geller M.J., Huchra J.P., Corwin H.G., 1994, *A.J.*, **108**, 437.
17. McGaugh S.S., 1994, *Nature*, **367**, 538.
18. Metcalfe N., Shanks T., Fong R.F., Roche N., 1994, *MNRAS*, submitted.
19. Metcalfe N., Fong R.F., Shanks T., 1994, *MNRAS*, submitted.
20. Peterson B.A., Ellis R.S., Efstathiou G.P., Shanks T., Bean A.J., Fong R., Zen Long Z., 1986, *MNRAS*, **221**, 233.